\documentclass{webofc} 
\usepackage[varg]{txfonts} 
\usepackage[english]{babel}
\usepackage[T1]{fontenc}
\usepackage[colorlinks=true,linkcolor=blue,citecolor=blue,urlcolor=blue]{hyperref}
\usepackage{ltablex}
\usepackage{booktabs}
\usepackage{multirow}
\usepackage{amsmath}
\usepackage{bm}
\usepackage[capitalize]{cleveref}

\usepackage{indentfirst}

\begin{document}

\title{Exotic atoms at extremely high magnetic fields:\\ the case of neutron star atmosphere}

\author{\firstname{Andrea} \lastname{Fontana}\inst{1}\fnsep\thanks{\email{andrea.fontana@pv.infn.it}} 
\and    \firstname{Alessandro} \lastname{Colombi}\inst{2}\fnsep 
\and    \firstname{Pietro} \lastname{Carretta}\inst{2}\fnsep
\and    \firstname{Alessandro} \lastname{Drago}\inst{3}\fnsep
\and    \firstname{Paolo} \lastname{Esposito}\inst{4}\fnsep
\and    \firstname{Paola} \lastname{Gianotti}\inst{5}\fnsep
\and    \firstname{Carlotta} \lastname{Giusti}\inst{1,2}\fnsep
\and    \firstname{Diego} \lastname{Lonardoni}\inst{6}\fnsep
\and    \firstname{Alessandro} \lastname{Lovato}\inst{7}\fnsep
\and    \firstname{Vincenzo} \lastname{Lucherini}\inst{5}\fnsep
\and    \firstname{Francesco} \lastname{Pederiva}\inst{7}\fnsep
}

\institute{INFN, Sezione di Pavia
\and       Dipartimento di Fisica, Universit\`a degli Studi di Pavia
\and       Dipartimento di Fisica e Scienze della Terra, Universit\`a degli Studi di Ferrara and INFN, Sezione di Ferrara
\and       Anton Pannekoek Institute for Astronomy, University of Amsterdam
\and       INFN - Laboratori Nazionali di Frascati
\and       National Superconducting Cyclotron Laboratory, Michigan State University and Theoretical Division, Los Alamos National Laboratory
\and       Dipartimento di Fisica, Universit\`a degli Studi di Trento and INFN, Trento Institute for Fundamental Physics and Application
}

\abstract{
The presence of exotic states of matter in neutron stars (NSs) is currently an open issue in physics. The appearance of muons, kaons, hyperons, and other exotic particles in the inner regions of the NS, favored by energetic considerations, is considered to be an effective mechanism to soften the equation of state (EoS). In the so-called \emph{two-families scenario}, the softening of the EoS allows for NSs characterized by very small radii, which become unstable and convert into a quark stars (QSs). In the process of conversion of a NS into a QS material can be ablated by neutrinos from the surface of the star. Not only neutron-rich nuclei, but also more exotic material, such as hypernuclei or deconfined quarks, could be ejected into the atmosphere. In the NS atmosphere, atoms like H, He, and C should exist, and attempts to model the NS thermal emission taking into account their presence, with spectra modified by the extreme magnetic fields, have been done. However, exotic atoms, like muonic hydrogen $(p\,\mu^-)$ or the so-called \emph{Sigmium} $(\Sigma^+\,e^-)$, could also be present during the conversion process or in its immediate aftermath. At present, analytical expressions of the wave functions and eigenvalues for these atoms have been calculated only for H. In this work, we extend the existing solutions and parametrizations to the exotic atoms $(p\,\mu^-)$ and $(\Sigma^+\,e^-)$, making some predictions on possible transitions. Their detection in the spectra of NS would provide experimental evidence for the existence of hyperons in the interior of these stars.
}

\maketitle

\section{Introduction}
\label{intro}
The presence of exotic states of matter in the core of neutron stars (NSs) is currently an open problem in physics, and great effort is devoted to its theoretical and experimental investigation~\cite{Glendenning:2012, Reisenegger:2016, Weber:2014}. The appearance of kaons, hyperons, and other exotic particles is likely to happen in the inner regions of a NS due to energetic considerations, and it offers an effective mechanism to soften the equation of state (EoS). This softening affects the entire structure of the star, reducing the pressure and therefore the maximum mass that the star can stably support. The observation of two NSs with masses as high as $2\,M_\odot$~\cite{Demorest:2010, Antoniadis:2013} seems however to exclude the presence of exotic particles in the core of the star, based on what we know about the interaction between these particles and normal nucleons. This apparent inconsistency, usually referred to as \emph{hyperon puzzle}, is due to the poor knowledge of the interactions involved in these exotic states of matter and to the difficulty of obtaining clear evidence of their presence in NSs from astrophysical observations.

In this work we propose a novel approach to attempt the identification of hyperons and other exotic particles in NSs: the search for signals of the possible formation of exotic atoms in the atmosphere of NS in the process of conversion into a quark star (QS) through the spectroscopic study of their radiation emission. In particular, the spectroscopy of exotic atoms, made in the simplest case by two Fermions like $(p\,\mu^-)$ (muonic hydrogen) or $(\Sigma^+ e^-)$ (Sigmium), can be inferred by the results obtained in the past for hydrogen or hydrogen-like atoms in the extremely strong magnetic field of NSs~\cite{CK:1972, Lai:2001}. It is expected that, under the condition of a strong magnetic field $\bm{B}$, atoms are of cylindrical shape, and that the traditional level structure observed in terrestrial experiments (\textit{gross}, \textit{fine}, \textit{hyperfine}) is superseded by a much simpler structure of Landau levels with principal quantum number $n$, with two additional quantum numbers for each level: $m$, which corresponds to the angular momentum projection on the field axis, and $\nu$, which quantizes the motion along $\bm{B}$. The evaluation of hypothetical new spectral lines, corresponding to transitions among these levels, could offer alternative interpretations of the anomalies seen in the existing data and/or suggest new observational strategies.

\section{Synopsis of neutron stars theory}
\label{sec:1}
Neutron stars are very peculiar astrophysical objects. Among their many interesting features, two are relevant for this work: the very high density and the extremely intense magnetic field. Considering nuclear saturation density $\rho_0 \approx 2.5\times 10^{14}\,\rm g/cm^3$, the density of matter in a NS varies from $\approx 3\times 10^{-3}\rho_0$ on the surface to $\approx 7\,\rho_0$ in the core, and the magnetic field ranges from $10^7 \,\rm G$ on millisecond pulsars to $10^{16}\,\rm G$ on magnetars, with an average value of $\approx 10^{12} \,\rm G$ on most NSs. 
\begin{figure}[!hbt]
	\centering
	\includegraphics[width=0.45\linewidth]{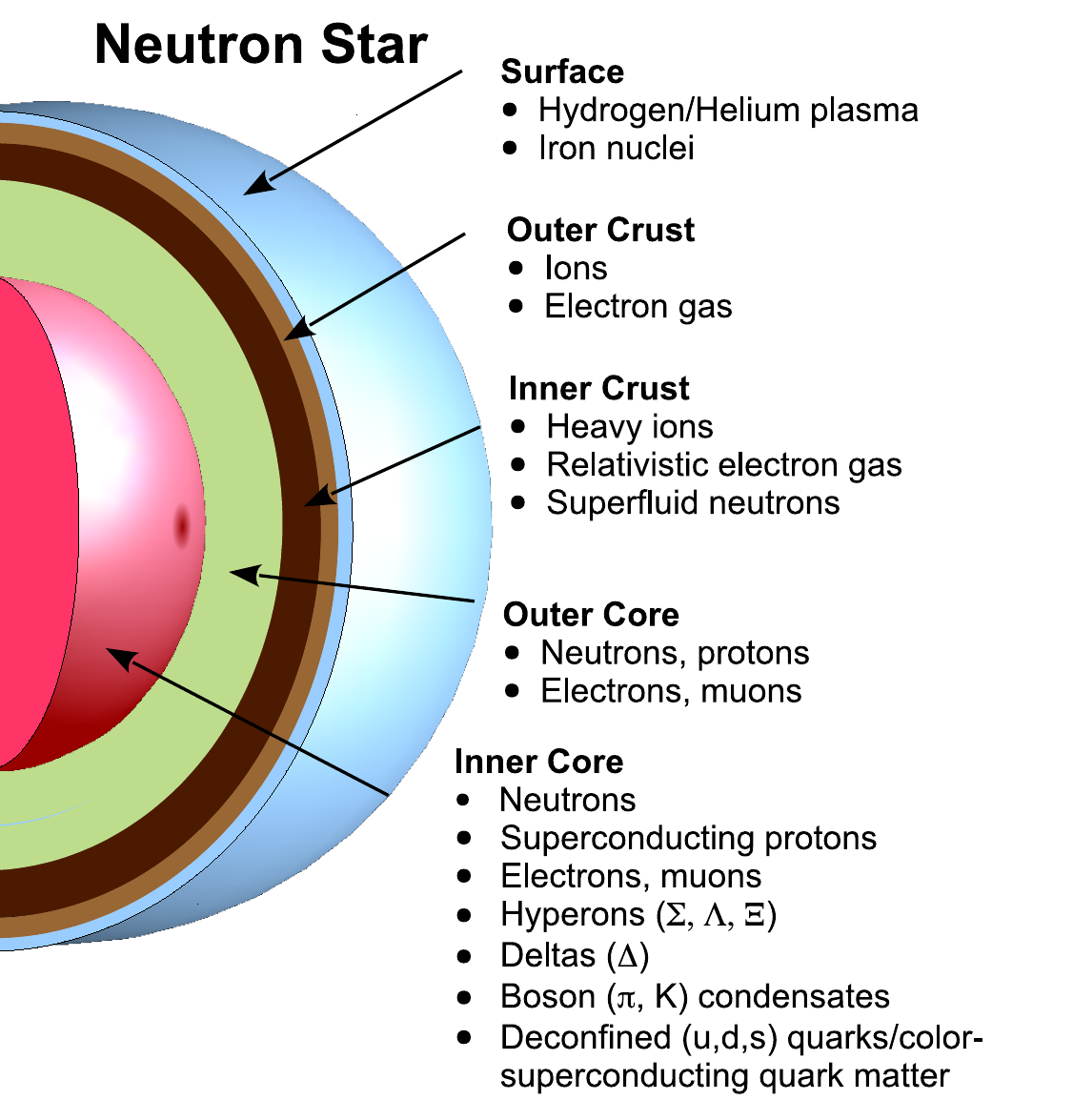}
	\includegraphics[width=0.52\linewidth]{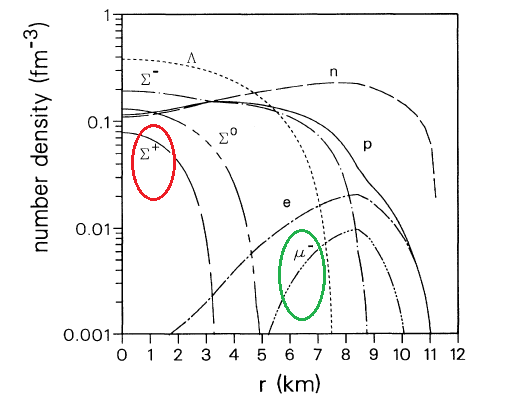}
	\caption[]{Structure of neutron stars~\cite{Weber:2014} (left) and radial distribution of the different particle fractions~\cite{Glendenning:2012} (right). Particles of interest for this study, $\Sigma^+$ and $\mu^-$, are highlighted in red and green, respectively.}
	\label{fig:ns}
\end{figure}
The extremely high density of NS matter increases towards the center of the star, and it spans more than 10 orders of magnitudes in the range of a dozen kilometers. This fact opens the possibility of extreme physical conditions, where different equilibrium situations are reached as the density increases, from neutronization and hyperon production, to pion (and possibly kaon) condensation, and even to quark deconfinement. Despite the name, NSs indeed contain a rich variety of subatomic particles whose equilibrium is governed by the chemical potential. In general, the baryon chemical potential increases with the density, and the energy of the system is lowered by sharing the conserved baryon number with other baryon species when their chemical potentials exceed their threshold values (i.e. their masses in vacuum). Thus, it is expected that super dense matter is populated by many baryon species, possibly even quarks.  The structure and composition of the star and the radial fraction of different baryonic and leptonic species in a NS for a given EoS is shown in~\cref{fig:ns}. As one can see, there are regions in the NS where different particles (and in particular hyperons) exist and overlap, opening the possibility for their mutual interaction. The reason why hyperons are expected to appear in the interior of NSs is very simple. Let us first we consider a matter made of neutrons only (pure neutron matter). As a first approximation, this system can be considered to be a gas of Fermions, with single-particle states occupied up to the Fermi energy, which is determined by the density. If the Fermi energy or, in the most general case, the chemical potential becomes equal to 
the difference in mass between a neutron and an electrically neutral hyperon, it is energetically convenient for the system to transform a neutron into a $\Lambda$~particle. The hyperon, being distinguishable from nucleons and not being involved in the Pauli blocking mechanism, will reduce the total energy of the system, leading to a reduction of the exerted pressure, and thus to a reduced maximum mass. Note that this simple argument does not take into account the effects of the interaction. If the potential between hyperons and nucleons is soft, one can expect that this description is still appropriate. However, in the presence of a strong repulsion, the energetic cost of creating hyperons might become such as to suppress their creation. This is now generally accepted as one of the possible explanations of the hyperon puzzle~\cite{Lonardoni:2015}.

\begin{figure}[!hbt]
	\centering
	\includegraphics[width=0.41\linewidth]{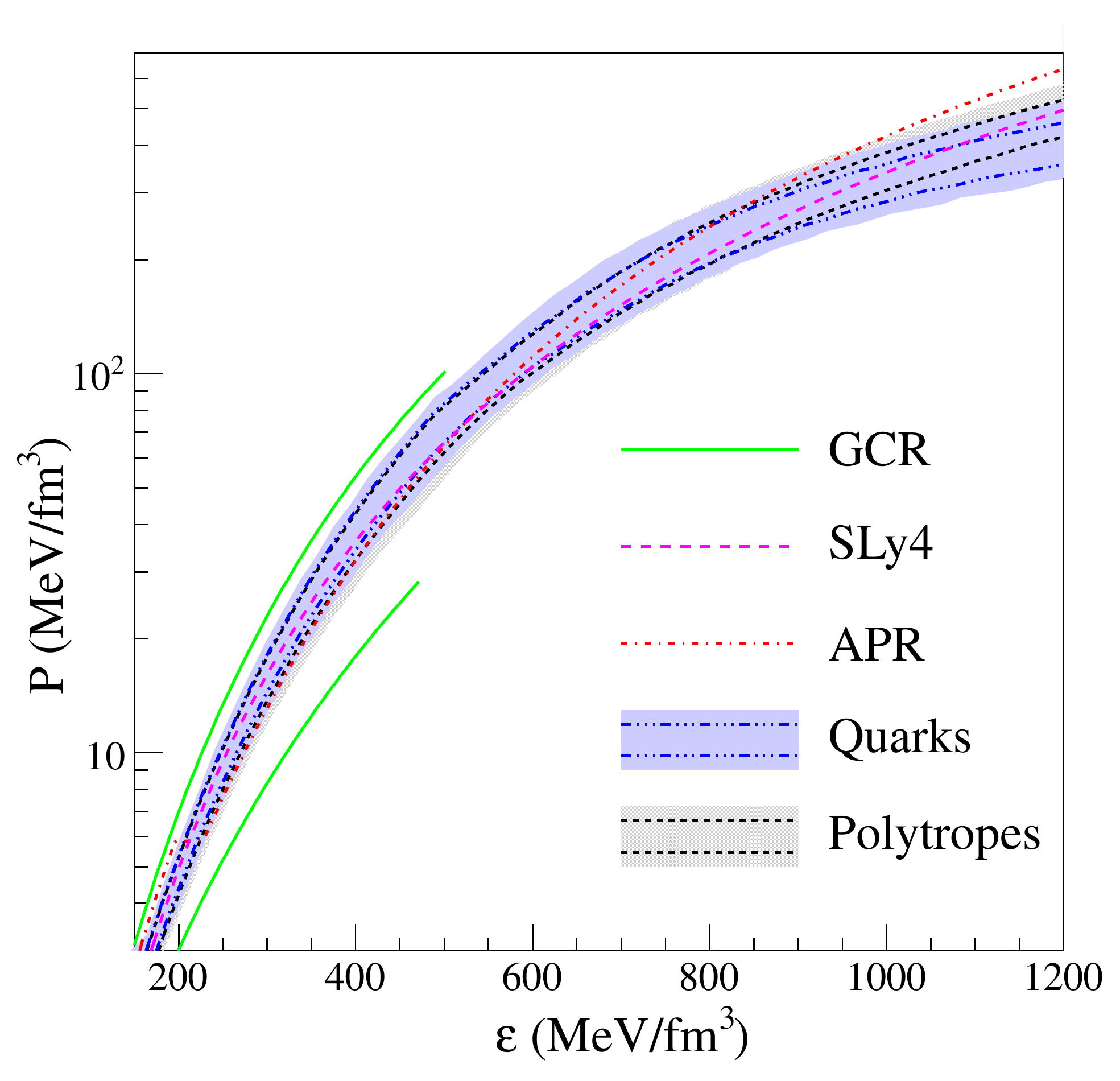}
    \qquad
	\includegraphics[width=0.41\linewidth]{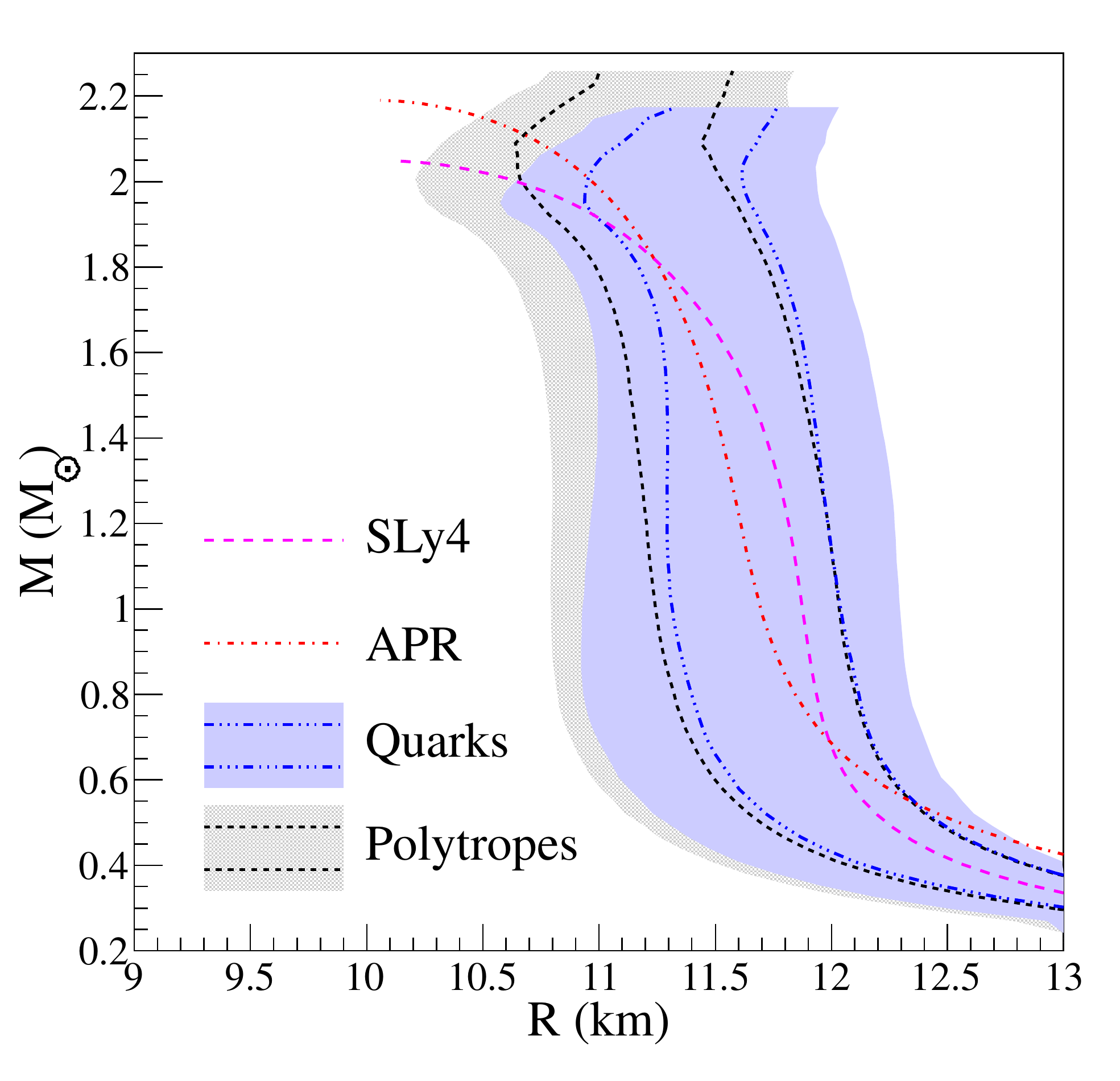}
    \caption{Constraints on the pressure vs. energy density function (left) and mass vs. radius relation (right) as deduced from astrophysical observations, according to the work of Steiner and Gandolfi~\cite{Steiner:2012}. The dashed and dot-dashed lines represent the 68\% confidence levels, the shadowed areas the 95\% confidence level. Curves represent the same quantities from the EoS evaluated in different schemes: GCR - quantum Monte Carlo~\cite{Gandolfi:2012}; SLy4 - Skyrme density functional~\cite{Chabanat:1995}; APR - variational calculation~\cite{Akmal:1998}.}
	\label{fig:eos}
\end{figure}

Different hypotheses currently exist on the internal structure of a NS, and the most accepted models foresee two typical structures: the so called \emph{traditional neutron stars} and the \emph{quark stars}~\cite{Weber:2014}. A third view is also proposed, the \emph{hybrid neutron stars}, as shown in Ref.~\cite{Glendenning:2012}. Many EoS have been proposed to describe these hypotheses, from the simple non interacting neutron gas, to more advanced parametrizations that take into account nuclear interactions and causality constraints. The situation is summarized in~\cref{fig:eos}, where the pressure $P$ in the star is plotted as a function of the energy density~$\varepsilon$. The density behavior of the EoS depends on the derivative of $P$ vs $\varepsilon$, and it is generally referred to as the \emph{stiffness} or \emph{softness} of the EoS. The stiffness is directly related to the maximum mass allowable for a given star, and it is therefore connected with astrophysical observations. Typical NS radii are of the order of $11-12\,\rm km$, and the average mass is $\approx 1.4\, M_\odot$, but fluctuations around this value are known~\cite{Kiziltan:2013}. Mass and radius are the main observables in the study of NSs: masses are measured with great precision from binary radio pulsars timing, but radius measurements are much more challenging, and no accurate data currently exist. A new set of constraints could follow from the recent observation of gravitational waves from a binary NS merger, particularly thanks to the analysis of the tidal deformability parameter from the inspiral signal. However, techniques to simultaneously infer both mass and radius have been proposed~\cite{Kramer:2009}, but not yet successfully applied. Therefore, based on the available information on NS mass and radii, it is currently difficult to discriminate between different EoS models.

\section{Neutron star/quark star conversion process}

\begin{figure}[hbt]
	\centering
	\includegraphics[width=0.8\linewidth]{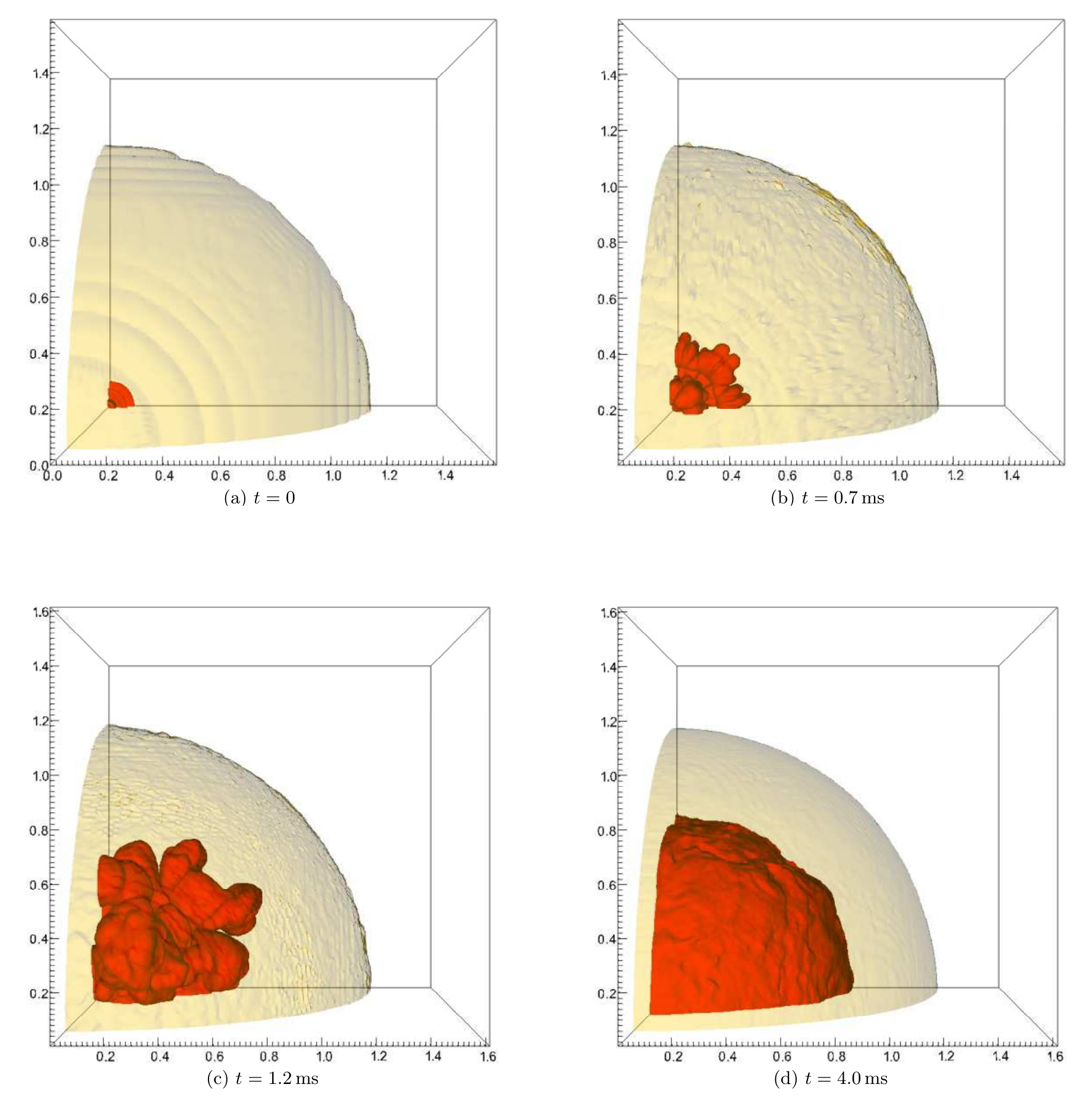}
	\caption[]{NS/QS conversion. Conversion front (red) and surface of the NS (yellow) at different times $t$: (a) $t=0$; (b) $t=0.7\,\rm ms$; (c) $t=1.2\,\rm ms$; (d) $t=4.0\,\rm ms$. Spatial units: $10^6\,\rm cm$~\cite{Pagliara:2013}.}
	\label{fig:massrad}
\end{figure}

In the so-called \emph{two-families scenario}~\cite{Drago:2016_2}, in which NSs co-exist with QSs, NSs can be very compact and have a maximum mass of about $1.5-1.6\,M_\odot$, while QSs can have large radii and be very massive, up to $2.75\,M_\odot$~\cite{Drago:2014,Drago:2016_1,Drago:2016_2}. Concerning the NS branch, with the increase of the central density, hyperons (and/or $\Delta$ resonances) will be produced. The increasing softening of the EoS allows the stars in this branch to reach very small radii, even significantly smaller than $11\,\rm km$ (providing a possible strong signature of this scenario). When the strangeness content at the center of the star (most likely due to a non marginal presence of hyperons) reaches a critical value, the star becomes unstable to nucleation of droplets of deconfined strange quark matter. At this point the NS converts into a QS (see~\cref{fig:massrad}).
The process of conversion of a NS into a QS has been studied in detail and within different approaches in a series of papers~\cite{Olinto:1987,Bombaci:2000,Bombaci:2016,Drago:2007,Herzog:2011,Pagliara:2013,Drago:2015,Keranen:2005}. The main result of these analyses is that the combustion proceeds in two phases: a first rapid burning, accelerated by hydro-dynamical instabilities, lasting only a few milliseconds and converting the core of the star, followed by a much slower combustion driven by diffusion of strange quarks from the burnt into the unburnt region, lasting few tens of seconds. During this second phase, material can be ablated by neutrinos from the surface of the star~\cite{Drago:2016_3}. This is particularly interesting, because it opens the possibility of ejecting in the atmosphere not only neutron-rich nuclei, but also more exotic material, such as hyperons. These particles could interact with the surrounding electrons or protons to form atomic structures, with recombination processes at rest similar to those occurring in laboratory with trapped particles~\cite{Amsler:2014}, or with in flight capture~\cite{Adeva:2017}.

Another possible process for generating exotic states of matter in the outer envelope of NSs could be the merger of two QSs, during which nucleosynthesis $r$-processes are believed to re-ignite, and exotic material could also be ejected and be involved in the process. This topic is of great interest nowadays after the recent multi-messenger observation of a kilonova event in the galaxy NGC 4993~\cite{Savchenko:2017}, in conjunction with the detection of gravitational waves GW170817 from a binary NS merger with the LIGO/VIRGO observatory~\cite{Abbott:2017}. 

\section{Exotic atoms on neutron stars}

The formation of exotic structures at the atomic level in the events outlined in the previous section is an intriguing problem that deserves to be investigated theoretically. In fact, given the uncertainties on the EoS, these atoms could provide new observables that enable to distinguish between the two types of compact stars, NSs and QSs.

\begin{figure}[!htb]
	\centering
	\begin{tabular}{c|c}
	\includegraphics[width=0.60\linewidth]{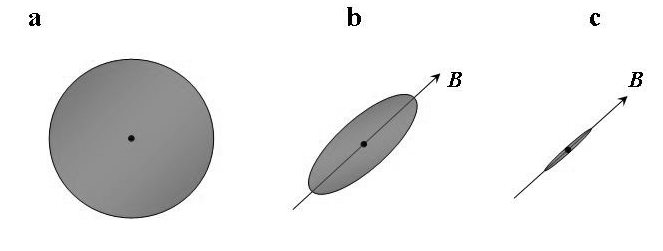} &
	\includegraphics[width=0.25\linewidth]{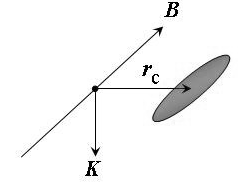}
	\end{tabular}
	\caption[]{Effects of a strong magnetic field on H atoms: a) $B \ll 10^9\,\rm G$, b) $B = 10^9\,\rm G$, c) $B = 10^{12}\,\rm G$ (gray areas: ellipsoids with probability to find an electron $P>e^{-1}$; solid dots: protons) (left). Decentered state, with an H atom confined by the magnetic field, but with the proton external to the electron cloud~\cite{Haensel:2007} (right).} 
	\label{fig:h01}
\end{figure}

It is widely accepted that regular atomic structures exist in the outer layers of the star, i.e. in the thin atmosphere. Indeed, results of the spectroscopy of atomic structures in NSs are present in the literature. For instance, hydrogen was investigated in detail after the discovery of pulsars~\cite{Smith:1972}, and observational results were obtained in the case of Fe lines in the X-ray spectra of accreting NSs~\cite{Ruder:1981,Lai:2001,Mori:2002}. Theoretical models for H, He, and C atmospheres have been recently proposed, and good fits of observed spectra with C models have been obtained, as in the case of the NS at the center of the supernova remnant Cassiopeia A with the Chandra X-ray observatory~\cite{Potatm:2009}. In addition, the hypothesis for the existence of exotic Coulombic systems, i.e. molecular ions such as ($\alpha pe$), ($\alpha \alpha e$), or (Li$^{3+}$Li$^{3+}e$), has been also investigated~\cite{Turbiner:2007}. These atoms, when immersed in the extreme magnetic fields found in the atmosphere of NSs, change dramatically their structure and behavior. Since the magnetic field is so strong that the Coulomb force can be treated as a perturbation, the atoms assume an elongated, cylindrical shape in the direction of the magnetic field (\cref{fig:h01} (left))~\cite{Meszaros:1992}, and the energy spectrum is much different from the zero---or low---field case, where the traditional quantum numbers $n$, $l$, and $m$ are appropriate to describe the spherically symmetric atom. In addition, due to the magnetic confinement, the so-called \textit{decentered} states, shown in~\cref{fig:h01} (right), are possible. These are rather unique bound systems, with the source of the field external to the electron cloud and kept together by the magnetic field. The energy levels are distributed according to the quantized cyclotron scheme of Landau levels with quantum number $n$, and, for each $n$, additionally described by two new quantum numbers: $m$, that quantizes the radial wave functions and it is related to the $z$-component of the angular momentum $l_z$, and $\nu$, the number of nodes of the longitudinal wave function.

\begin{figure}[!htb]
	\centering
	\includegraphics[width=0.56\linewidth]{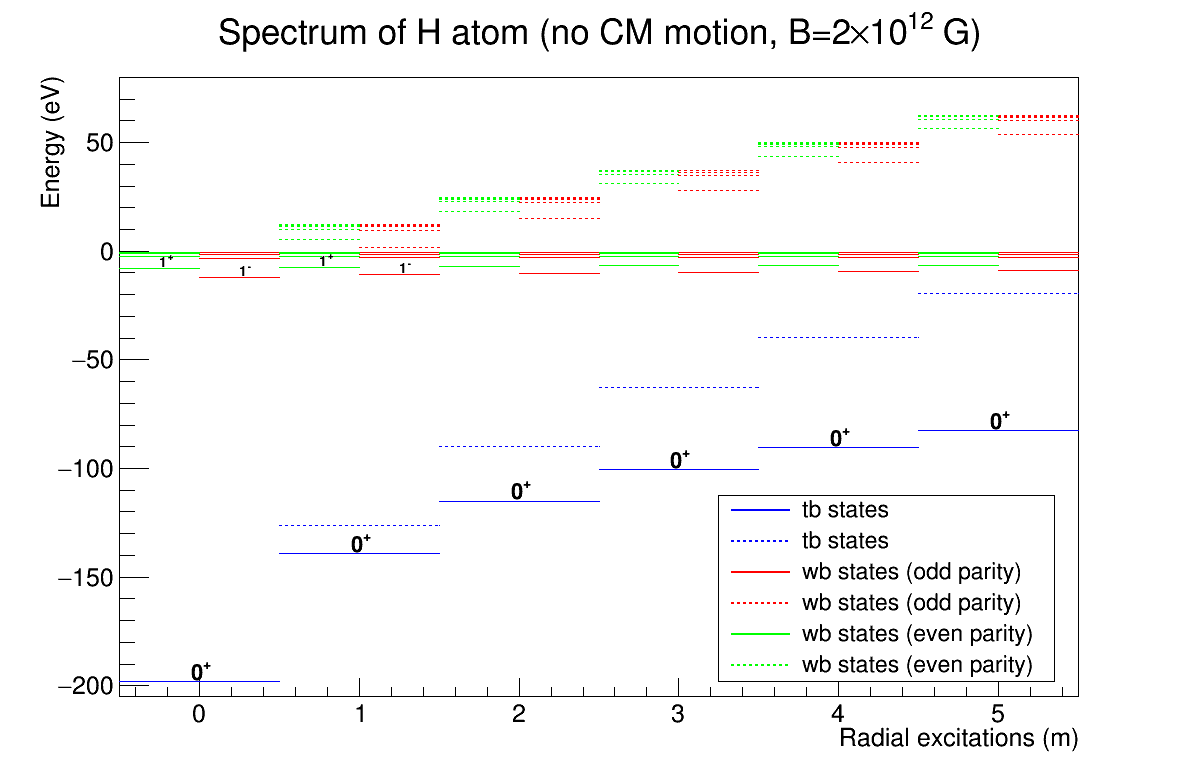}
	\includegraphics[width=0.43\linewidth]{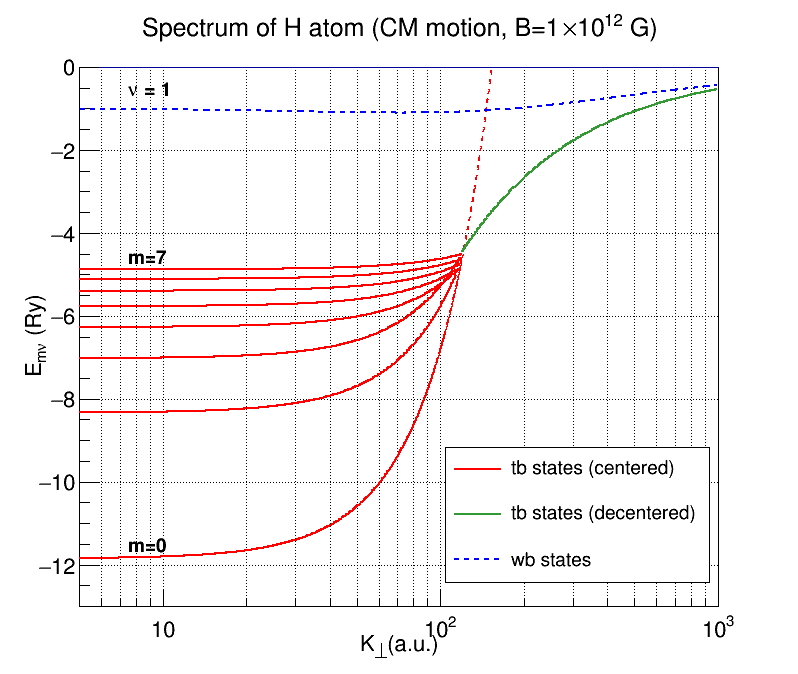}
	\caption[]{The hydrogen atom energy level structure in a strong magnetic field $B=2\times10^{12}\,\rm G$. Static spectrum with infinite proton mass (left, continuous lines) and with finite proton mass (left, dashed lines). Dependence of the lowest energy levels on the transverse pseudomomentum describing the center of mass motion (right).}
	\label{fig:h02}
\end{figure}

The expected modification of the H energy levels for a field $B=2\times10^{12}\,\rm G$ is shown in~\cref{fig:h02} (left). These states are obtained by solving the one-dimensional Schr\"{o}dinger equation for the wave functions $f_{m \nu}$:
\begin{align}
- \frac {\hbar ^2}{2 m_e \rho _0 ^2} f'' _{m \nu} - \frac {e}{\rho _0} V_m (z) f _{m \nu} = E _{m \nu} f _{m \nu} \qquad m, \nu =0,1,2,\dots\,,
\end{align}
where the average potential $V_m(z)$ is derived from the Landau wave functions radial part (see~\cite{Lai:2001} for details). The eigenvalues of this equation are classified according the quantum numbers $m$ and $\nu$ and to the parameter $b=B/B_0$\footnote{$B_0$ is a reference field typical for NSs ($B_0 = m _e ^2 e^3 c\,\hbar^{-3} = 2.3505 \times 10^9\,\rm G$), corresponding to the situation: cyclotron radius $\equiv$ Bohr radius.} in \textit{tightly-bound} (tb) states ($m \ge 0$, $\nu=0$) and \textit{weekly-bound} (wb) states ($m \ge 0$, $\nu > 0$). The latter present a doublet sub-structure according to the parity of the solution of the one-dimensional Schr\"{o}dinger equation.
With the definitions $\rho_0=\Big( \frac{\hslash c}{2 e B} \Big)^{1/2}$ and $\rho_m=(2m+1)^{1/2} \rho_0$, the energy levels are given by the following expressions for the tightly-bound states:
\begin{align}
E_{m0} \approx - 0.16\,A\, l _m ^2 \quad \text {a.u.} \qquad (\text {for } 2m +1 \ll b )\,,
\end{align}
where: 
\begin{align}
A &=
\begin{cases}
1+1.36\times 10^{-2}\Big[\ln (1000/b) \Big]^{2.5} & b<10^3 \\ \nonumber
1+1.07\times 10^{-2}\Big[\ln (b/1000) \Big]^{1.6} & b\ge 10^3
\end {cases}\,,\\
l _m  &= \ln \left(\frac{b}{2 m +1}\right)\,, 
\end{align}
and for the weakly-bound states:
\begin{align}
E _{m \nu} = -\frac {1}{(2 \nu _1 + \delta)^2} \quad \text {a.u.}\,,  
\end{align}
where:
\begin{align}
\delta &=
\begin {cases}
2 \rho / a_0 \qquad &\nu = 2 \nu _1 -1 \\ 
\Big[\ln (a_0/\rho _m)\Big]^{-1} \qquad &\nu = 2 \nu_1 
\end {cases} \,.
\end{align}

The effect of the finite proton mass, due to the replacement of the electron mass with the reduced mass, introduces a small correction to each level given by $\Delta E = 29.6 \, m \, \frac{B}{4.7\times10^{12}\,G} \,\rm eV$. The center-of-mass motion introduces an additional degree of freedom, the \textit{pseudomomentum} $\bm K$~\cite{Lai:2001}, which is derived from the canonical momentum $\bm \Pi$:
\begin{align}
\bm K = \bm \Pi - \frac {e}{c} \bm B \times \bm r \,,
\end{align}
and it is used to define the position of the guiding center ${\bm r}_c$:
\begin{align}
{\bm r}_c = \frac {c {\bm B} \times {\bm K}_{\perp}}{e B ^2}\,,
\end{align}
where $ \bm K _{\perp}$ is the pseudomomentum component transverse to the magnetic field direction. In terms of these quantities, the energy levels for the deeply-bound states can be classified as two different solutions, the centered and and the decentered states with energies:
\begin{align}
\text{centered:\phantom{de}}\quad E _{m 0} (K_{\perp}) &\approx E _m + \frac {K_{\perp} ^2}{2 M_{\perp\,m} } \,, \nonumber \\
\text{decentered:}\quad E _{m 0}(K_{\perp}) &\approx -\frac{b}{K_{\perp}}\,,
\end{align}
where $E_m$ is the energy of a bound electron in the fixed Coulomb potential, $M_{\perp\,m}$ is the effective transverse mass, and all quantities are expressed 
in atomic units (see~\cite{Lai:2001} for details). Similar parametrizations are also found for the weakly-bound states, resulting in the typical spectrum shown in~\cref{fig:h02}~(right).

In a typical spectroscopic study of these atoms the wavelength associated to a transition between these levels can be estimated by using the equations above and by taking into account the usual selection rules~\cite{Smith:1972}. For hydrogen, for instance, the $\nu=1^-\to \nu=0^+$ transition (for $m=0$) produces radiation in the extreme UV regions ($\sim 160\,\rm eV$), but additional transitions are observed for other light atoms (H, He, and C) in the atmospheres of NSs. Moreover, spectral features in the X-ray thermal emission of some isolated pulsars have been observed. While their origin is debated, their interpretation as atomic transitions between different $m$ levels in a magnetized atmosphere is not ruled out~\cite{Ruder:1994,Kerkwijk:2007,Borghese:2017}.

\begin{figure}[b]
	\centering
	\includegraphics[width=0.47\linewidth]{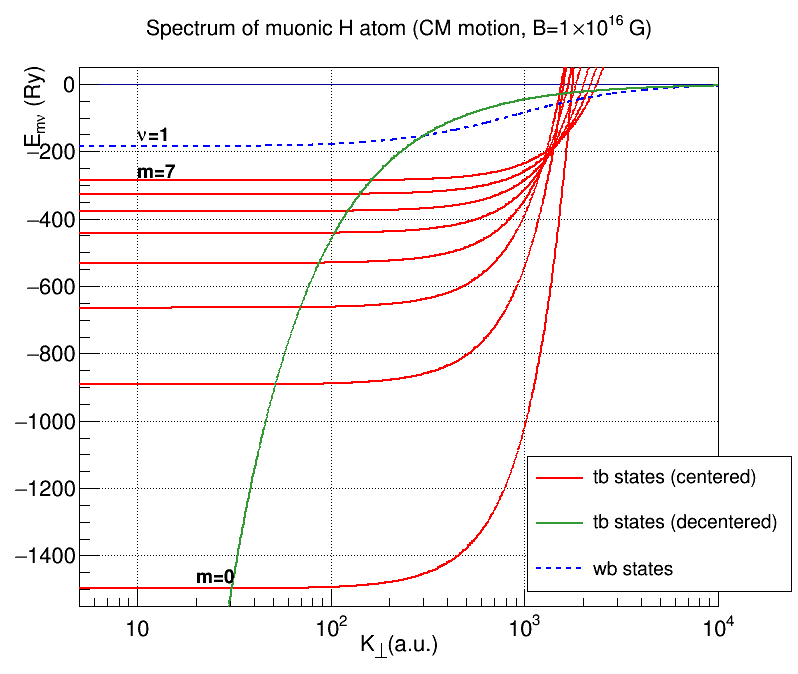}
    \qquad
	\includegraphics[width=0.47\linewidth]{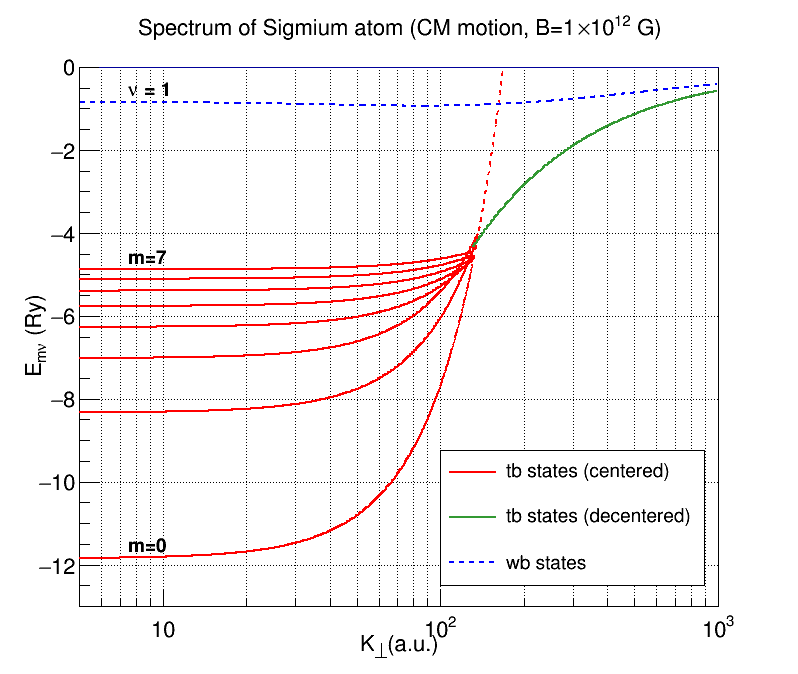}
	\caption[]{Lowest energy levels for deeply bound states (centered and decentered) and weakly bound states obtained from the hydrogen calculations for the hypothetical muonic hydrogen (left) and Sigmium (right) atoms.}
	\label{fig:musig}
\end{figure}

It is important to note that, from the theoretical point of view, these results are valid for $2m +1 \ll b$ and $100 \le b \le 10^6$~\cite{Lai:2001}. The latter interval, corresponding to $B_{\min} \approx 3 \times 10^{11}\,\rm G$ and $B_{\max} \approx 3 \times 10^{15}\,\rm G$, provides a good coverage of the magnetic field distribution typical of NSs. To generalize these studies to exotic atoms like muonic hydrogen $(p\,\mu^-)$ or Sigmium $(\Sigma^+\,e^-)$, in first approximation it is possible to rescale the reduced mass and to redefine the parameter $b$ in the equations found for hydrogen. Since the present approach is valid for $100 \leq b \leq 10^6$, one has to consider that a change in the reduced mass would imply a change in the magnetic field range where the solutions are still valid. For example, for muonic hydrogen, for which we have $b = \frac{\hbar^{3} B }{m_\mu ^2 e^3 c}$, with a ratio $\frac{b}{b^H} \sim 10^{-2}$, an increase of the magnetic field of $ 10^{16} \,\rm G$ is required, while for Sigmium, for which $b = \frac{\hbar^{3} B }{m_\Sigma ^2 e^3 c}$, with a ratio $\frac{b}{b^H} \sim 1$, no changes in the magnetic field with respect to the hydrogen case are required. The corresponding spectra with the center of mass motion correction are shown in~\cref{fig:musig}.
The full analytical expressions of the wave functions and eigenvalues of the above described levels cannot be easily calculated for exotic atoms. The use of analytical approximations, as shown for example in~\cite{Potekhin:1998}, or ab-initio numerical approaches to solve the Schr\"{o}dinger equation~\cite{Mori:2002}, are both possible strategies that require further investigation. However, as an indication of the expected spectral lines, we present in~\cref{tab:uno} an example of possible transitions among these hypothetical states obtained from the above equations. To assess the observability of the calculated lines, the knowledge of the formation rate of 
these systems in the atmosphere of a NS, either at the end of the conversion process into a QS, or in the merger with another NS, is essential. Therefore, future studies on this topic would require a dynamical simulation of the NS evolution and they could include results obtained from other theoretical studies on hypernuclear physics, on exotic atoms formation, and on atmospheric models of NSs~\cite{Potekhin:2014}. 
\begin{table}[!htb]
\centering
\begin{tabular}{ccccc}
\toprule
Atom  & $K_{\perp}$\,(a.u.) & Energy\,(eV) & $\lambda$\,(nm) & Spectral region \\
\midrule
\multirow{4}{*}{Hydrogen} &0 &147.5877 &8.4 & \multirow{4}{*}{UV}\\
  &10  &146.6178 &8.5 &  \\
  &50  &129.1634 &9.6 &  \\
  &100 &77.6384  &16  &  \\
\midrule
\multirow{4}{*}{Muonic hydrogen} &0 &17848.5 &0.06946 & \multirow{4}{*}{X-ray}\\
  &10  &17851.5 &0.069453 &  \\
  &50  &17870.2 &0.069381 &  \\
  &100 &17885.1 &0.069323 &  \\
\midrule
\multirow{4}{*}{Sigmium} &0 & 149.832 & 8.27 & \multirow{4}{*}{UV}\\
  &10  & 149.029 & 8.32  &  \\
  &50  & 134.488 & 9.22  &  \\
  &100 & 91.7076 & 13.52 &  \\
\bottomrule
\end{tabular}
\caption{Examples of possible transitions for H, muonic hydrogen, and Sigmium for 
different values of $K_{\perp}$.}
\label{tab:uno}
\end{table}

\section{Conclusions}

Hydrogen-like atoms, involving $\mu^-$, $\Sigma^+$, or other exotic constituents could be formed in the NS/QS conversion. In this work we have investigated the energy level structure of these hypothetical exotic atoms, and made some predictions on transitions that could help to discriminate among different EoS. As a further step in this direction, it would be interesting to estimate the radial exotic fractions during the conversion process, and to numerically solve the one-dimensional Schr\"{o}dinger equation for exotic atoms. If the results are confirmed, the proposed calculations could give a direct evidence of the presence of hyperons in NSs. 

\section{Acknowledgments}
We thank I. Bombaci, A. Gal, D. Logoteta, A. Ramos, and I. Vida\~{n}a for insightful comments and suggestions. The work of D.L. was supported by the U.S. Department of Energy, Office of Science, Office of Nuclear Physics, under the FRIB Theory Alliance Grant Contract No. DE-SC0013617 titled ``FRIB Theory Center - A path for the science at FRIB'', and by the NUCLEI SciDAC program. P.E. acknowledges funding in the framework of the NWO Vidi award A.2320.0076.


\end{document}